# Some Aspects of Galactic Cosmic Ray Acceleration

Yousaf M. Butt

*Harvard-Smithsonian Center for Astrophysics, 60 Garden St., Cambridge, MA, USA*

**ABSTRACT**

**I give a synopsis of two aspects of the Galactic Cosmic Ray (GCR) acceleration problem: the importance of the medium energy $\gamma$-ray window, and several specific astrophysical sources which merit further investigation.**

## *INTRODUCTION*

Cosmic rays continually rain down on our heads, but we have virtually no better idea where they come from than did investigators fifty years ago (eg. Shklovskii, 1953). $\gamma$-ray astronomy has been the tool of choice to investigate their origin(s) since a local overdensity of freshly accelerated GCRs are expected to produce co-spatial high energy ($\gamma$-ray) photons. Of course, the proper interpretation of such $\gamma$-ray data (specifically, whether electrons or nuclei are generating the detected radiation) requires as broad a multiwavelength coverage as practical: *$\gamma$-ray data is necessary but not sufficient* to prove the existence of a GCR acceleration site. Unfortunately, sufficient multiwavelength coverage of requisite fidelity is often simply unavailable. As outlined by Francis Halzen in this session, now neutrino astronomy also appears poised to contribute to GCR research in the 'non-wavelength' window.

Despite almost 40 years of concerted effort in $\gamma$-ray astronomy, not a *single* source of nucleonic GCRs has yet been firmly identified. The main reason for this is the rather marginal spatial resolution of virtually all $\gamma$-ray instrumentation flown, which results in severe source confusion – eg. are the $\gamma$-rays coming from a Supernova Remnant (SNR) shock or a related, or unrelated, pulsar? The latest ground-based stereo Cherenkov telescopes and the next-generation orbiting GeV detectors promise to finally overcome this limitation.

Currently, there are two favored theoretical mechanisms for accelerating GCRs, and both invoke shocks: either the shocks of isolated SNRs in the $\sim 10^4$-$10^5$ yrs after explosion (eg. Drury et al., 2001; see Torres et al., 2003 for a recent review and Plaga, 2002 for an alternative hypothesis); or, the cumulative shocks of massive stars and/or multiple SNRs in massive stellar associations (or after evolution, 'superbubbles') which last $\sim 10^7$ yrs (eg. Cesarsky & Montmerle, 1983; Bykov 2001). It is certainly also possible that both mechanisms could be operating simultaneously in the Galaxy. Although as a class only SN appear to have the required power input to the ISM to explain the local energy density of GCRs, it is unclear whether the physical mechanism for accelerating nucleonic GCRs requires that multiple SNRs be embedded in a superbubble or stellar cluster. Interestingly, although the bulk of SN are expected to occur in superbubbles, many more individual, isolated, SNRs have been identified and cataloged than superbubbles.

## *THE IMPORTANCE OF THE MeV WINDOW*

Although the acceleration of high energy CRs in a given source cannot be proven without the detection of similarly high energy (GeV→TeV, and even higher energies) radiation localized with the source, the lower-energy MeV window is still very important in discriminating the nature of the particles generating the detected $\gamma$-rays: $e^-$'s *vs*. nuclei. This is because in the case of a hadronic origin there is a plateau expected in the $\gamma$-ray spectrum in the $\sim 1 \rightarrow 100$ MeV range. Such a discontinuous 'plateau-ing' of the photon spectrum is not expected from purely leptonic bremsstrahlung or inverse-Compton emissions. The inability to distinguish the nature of the particles generating the detected $\gamma$-rays has been a major stumbling block in nucleonic GCR origin studies (eg. see Reimer & Pohl, 2002 and Butt et al., 2002). High signal-to-noise data in the soft-$\gamma$ window greatly simplifies our ability to assign a hadronic *vs*. leptonic origin to the detected $\gamma$-rays (Fig. 1). This is a direct consequence of characteristic shape of the hadronic pion-decay spectrum: in fact, were it not for the existence of, and emission from, hadronically-generated secondary electrons, we would expect a highly distinctive, symmetrical pion 'hump' centered at $\sim 67.5$ MeV ($= \frac{1}{2}m_{\pi}$) in the case that hadrons were producing the γ-rays (eg. Schlickeiser 1982). Note that in Schlickeiser's paper a steady-state leptonic population was assumed so that the expected plateau in SNRs/stellar associations' hadronic emission will be even more distinctive than shown in that study as steady-state has not been reached in the typical lifetimes of those objects, and thus less secondaries are present.

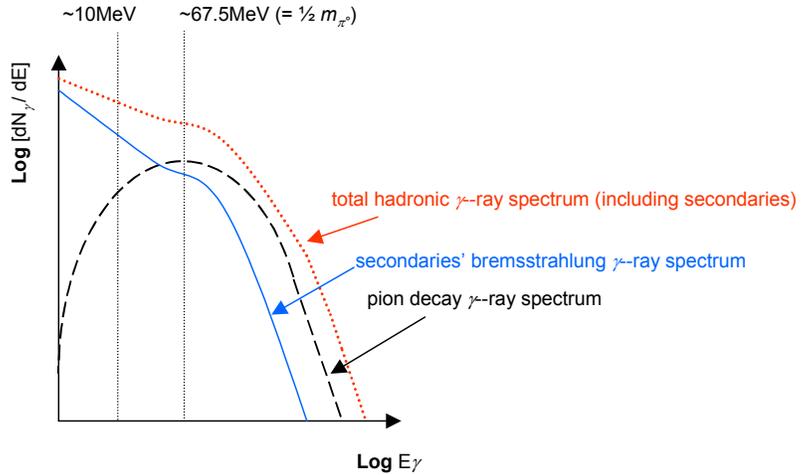

Fig.1: High signal-to-noise data in the ~1-100 MeV range are very useful in discriminating between a hadronic vs. leptonic origin of detected high-energy γ-rays. In the case of a hadronic origin, a plateau in the photon spectrum in the ~1-100 MeV range is expected due to the emission from secondary electrons generated in hadronic interactions. Such a discontinuous 'plateau-ing' is not expected from purely leptonic bremsstrahlung or inverse Compton emissions. Please see Schlickeiser (1982) for further details, but note that there a steady-state leptonic population was assumed so that the expected plateau in SNRs/stellar associations' hadronic emission will be even more distinctive than shown in that study as steady-state has *not* been reached in the typical lifetimes of those objects and thus less secondaries are present. [*Figure adapted from Schlickeiser (1982)*].

Though the INTEGRAL satellite covers the 1-10 MeV bandpass with relatively good spatial resolution, there is unfortunately a gap in the important ~10-100 MeV range. (The ≳50 MeV bandpass will be covered by AGILE and GLAST instruments in the near future.) Orbiting detectors, such as the proposed MEGA instrument (Kanbach et al., 2003) would thus be very useful for GCR studies by filling-in this bandpass gap.

## SPECIFIC SOURCES

Below I provide an *incomplete* list of sources which would be useful to observe in γ-rays as well as other wavelengths. I have divided the sources into three categories: isolated SNRs, stellar clusters/superbubbles, and extragalactic starburst galaxies. The list is not exhaustive and, certainly, there are many other sources worth investigating.

### *I. Isolated Galactic Supernova Remnants (SNRs)*
*a. G347.3-0.5 (RX J1713.7-3946)*

The SNR G347.3-0.5 (RX J1713.7-3946) and its environs are amongst the best laboratories for investigating shock-driven particle acceleration processes in our Galaxy. Photons spanning ~17 orders of magnitude in frequency, from radio through TeV range, have been reported as being plausibly associated with this SNR, or with its putative shock-molecular cloud interactions (Fig 2). Deeper radio observations of this source would help pin down the contributions of electrons *vs.* nucleons to the $\gamma$-ray emission. High resolution mm observations of 1→0, 2→1, and higher rotational level transitions of CO would help better determine the cloud excitation, and possibly, also whether the cloud shape 'meshes' with the SNR shock front. Similarly, high spatial resolution (CHANDRA) X-ray observations are needed towards the N of the remnant to image whether indeed the shock front and clouds are interacting. This is important in settling whether the clouds and the SNR are interacting, which is the key in determining the distance to the SNR. (The distance to the SNR is currently *assumed* be the same as the clouds with which it is *thought* to be interacting). Gamma-ray observations would also help narrowing down the rather large EGRET error box associated with this source. Specifically, one would like to know whether $\gamma$-ray emission is consistent with the location of the molecular clouds, and whether it's spectrum agrees with a hadronic origin. The recent report of a hard ASCA source (AX J1714.1-3912) seen immediately adjacent to the SNR rim (Uchiyama et al., 2002) and superposed with 'cloud A' lends particular urgency and significance to multiwavelength observations of this highly non-thermal SNR.

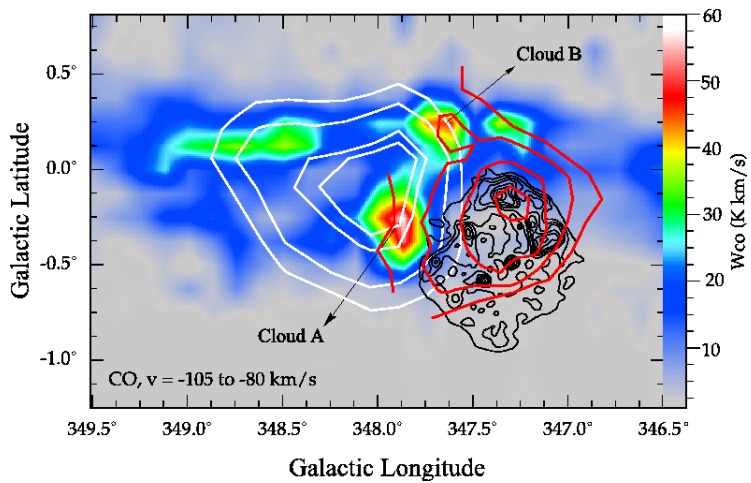

Fig. 2: An overlay map in Galactic coordinates showing SNR G347.3-0.5 in black X-ray intensity contours (ROSAT PSPC; Pfeffermann & Aschenbach, 1996). Red depicts the TeV significance contours (Enomoto et al., 2002). In white are the location probability contours of the GeV range EGRET source 3EG J1714-3857 (Hartman et al., 1999). The overall colorscale of the map indicates the intensity of the CO(J=1→0) emission, and, consequently, the column density of the ambient molecular clouds in the velocity interval $v_{lsr}$=-105 → -80 km/sec *thought* to be associated with the SNR, corresponding to a kinematic distance of 6.3±0.4 kpc (Slane et al., 1999; Butt et al., 2001). The hard X-ray source AX J1714.1-3912 is located towards 'cloud A' – see Uchiyama et al. (2002). JEM-X and IBIS data will be extremely useful to investigate whether there is hard X- & γ-ray emission from the cloud region. The spectrum of such radiation will provide a strong clue as to the nature of particles producing the γ-ray emission associated with this SNR. Figure from Butt et al. (2002).

### b. The SNRs W28 (G6.4-0.1); W44 (G34.7-0.4); and W66 ($\gamma$-Cygni, G78.2+2.1)
All three of these SNR-EGRET $\gamma$-ray source coincidences have been discussed extensively in the literature, and are reviewed in detail by Torres et al. (2003). These SNRs all appear to be in interaction with neighboring massive molecular clouds which may explain the detected $\gamma$-ray emissions; however, all three $\gamma$-ray sources are also in coincidence with energetic pulsars/compact objects which could, alternatively – or, in addition – be generating the $\gamma$-rays. Similar figures as Fig. 2 above for all three cases are available in Torres et al. (2003) [see Fig.'s 21, 22, 25, 28 therein], and are not reproduced here in the interest of brevity. Even before AGILE and GLAST are launched, INTEGRAL observations of these SNRs are needed to better localize the $\gamma$-ray emission region(s). Such data will also simultaneously greatly aid in clarifying the nature (hadronic *vs.* leptonic) of the emissions in these 3 important SNRs by providing a MeV spectrum (eg. see Fig. 1). Although not detected in the TeV range with the previous generation of Cherenkov telescopes (eg. Buckley et al., 1997), deeper observations with the updated arrays would be of great use.

### c. RCW 86 (MSH 14-63, G315.4-2.3)
RCW 86 is a bright shell-like SNR with *non-thermal* X-ray and radio emission which is especially strong towards the SW, where there is also evidence of interaction with a molecular cloud (eg. Borkowski et al., 2001; Rosado et al., 1996). In fact, the intensity of the the non-thermal X-ray emission from the SW shell is even brighter than that of SN 1006. RCW 86 also appears to be associated with a neighbouring OB association (Westerlund, 1969). Very recently the CANGAROO collaboration has reported the detection (at the ~$4\sigma$ level) of TeV range $\gamma$-rays from the non-thermal X-ray bright southwest shell at a level of ~$(3\pm0.7)\times10^{-12}$ photons cm$^{-2}$ sec$^{-1}$ (Watanabe et al., 2003). Vink et al. (2000) and Gvaramadze & Vikhlinin (2003) also report the presence of point-like X-ray sources towards the same SW shell. Data in the INTEGRAL waveband, and at higher energies, of this SNR will be useful to test whether there is associated MeV-range $\gamma$-ray emission from the SW shell; again, the spectrum of such putative emission will greatly aid in discriminating the origin of the high-energy flux. Due to their high spatial resolution, JEM-X (and IBIS) data will be especially useful in examining whether it is the point-like sources in the SW or rather the extended shock there that is generating the $\gamma$-rays.

### d. Monogem Ring SNR/PSR B0656+14
The Monogem Ring is a very large (~20° diameter), bright SNR with a young radio pulsar PSR B0656+14 projected virtually at its geometric center (see Fig. 1 in Thorsett et al., 2003 for an X-ray mosaic image). Very recently, Thorsett et al. (2003) have convincingly argued that the pulsar and the SNR are indeed related (contrary to some previous studies), and that this SNR – *by itself* – may be responsible for the existence of the 'knee' in the GCR spectrum at ~3 PeV ($3\times10^{15}$ eV). Indeed, Erlykin & Wolfendale (1997, 2003) already theorized that a single SNR of age ~100kyr and ~325pc from earth could by itself explain the knee feature – these parameters agree surprisingly well with those of the Monogem Ring SNR. In contrast to the previously mentioned SNRs in this presentation there is no high-energy EGRET source or reports of TeV emission associated with this SNR. However, this may not be surprising given its large size which would tend to 'wash-out' the signal due to the integrated background over the large SNR region.

## II. Galactic OB Associations/Superbubbles
### a. Cygnus OB2
Cyg OB2 is the most massive OB association known in the Galaxy, thought to contain in excess of 2500 OB type stars (Knoldseder 2002). Indeed, Hanson (2003) mentions several new massive star members in this association which were previously unknown due to the heavy extinction in the Cygnus direction. Recently, the HEGRA Chrenkov array group has reported a steady and extended TeV-range $\gamma$-ray source within this association (TeV J2032+4032; Aharonian et al., 2002). An EGRET source, 3EG 2033+4118, and a GeV source are coincident with, and nearby, the TeV source, respectively. We have carried out preliminary CHANDRA and VLA observations of this intriguing TeV source and have argued that the most likely explanation is that nuclei (rather than $e^-$'s) accelerated to high energies are responsible for the $\gamma$-ray emissions (Butt et al., 2003). If confirmed by future observations, such a scenario would strongly support a stellar association/superbubble origin of GCRs, as has been theoretically argued for some time (eg. Cesarsky & Montmerle, 1983; Bykov, 2001; and Parizot, 2002).

*b. RCW 38 & NGC3603*

I group these two stellar clusters together since recent CHANDRA observations have found evidence for hard, diffuse X-ray emission associated with both. Diffuse *non-thermal* X-ray emission from the southern stellar cluster RCW 38 was found in a recent ~100 ksec CHANDRA observation (Wolk et al., 2002). This is one of the brightest HII regions at radio wavelengths (eg. Wilson et al., 1970), and is a natural candidate for deeper multifrequency observations. The X-ray emission fills the center of a radio ring, reminiscent of some shell-type SNRs (Fig 3) and is very intriguing in the context of a young cluster/HII region. If hard X-ray/$\gamma$-ray emission is discovered superposed to the non-thermal X-rays, this would be a very significant result and would argue strongly in favor of collective shocks from the young stars accelerating GCRs to very high energies as proposed by, eg., Cesarsky & Montmerle (1983). (The age of the cluster is too young for there to be any 'contaminating' contributions from SNRs).

NGC 3603 is among the most massive and luminous visible starburst regions in the Galaxy. A recent 50 ksec CHANDRA observation has similarly found hard diffuse X-ray emission spatially coincident with the cluster core (Moffat et al., 2002). Those authors attribute this emission to the collective effects induced by the multiple, colliding stellar winds from the large population of massive stars in the region. As in the case of RCW38, if future observations find significant hard X-ray/$\gamma$-ray emission then this would be a major result since it would again favor the collective stellar wind hypothesis of the origin of (at least some fraction of) GCRs.

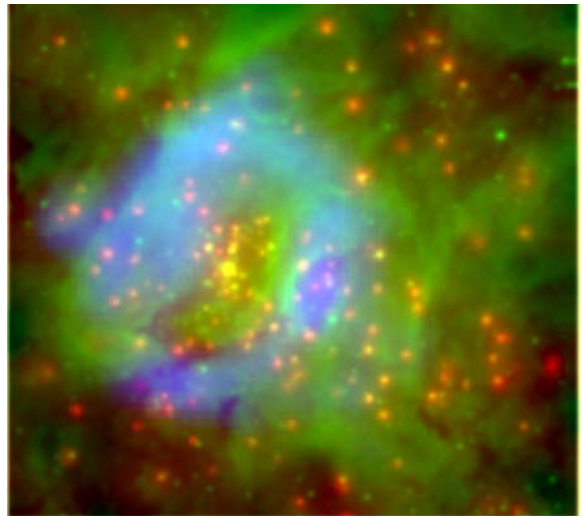

Fig. 3: This multiwavelength composite of the young star cluster RCW 38 shows the Chandra data (0.5-8 keV) in red, the VLT/ISAAC infrared data (K band 2.2 micron) in green and the ATCA radio data (1660.0 MHz, 18cm) in blue. The X-ray image was taken in December 2001, the infrared in November of 1998 and the radio in May of 1996. The radio 'shell' is found to surround the diffuse *non-thermal* X-ray emission centered at the core of the cluster. This morphology is reminiscent of some shell-type SNRs. Figure from Wolk et al., 2004 (in prep.) For an illustration of the diffuse non-thermal X-ray component alone please see Wolk et al. (2002).

CHANDRA has also detected diffuse X-ray emission from the Omega (M17/W38) & Rosette Nebulae (NGC2237-2246) but since the emission is soft (Townsley et al., 2003 & Dunne et al., 2003), we do not anticipate that these objects are significant GCR accelerators.

*III. The Starbursts NGC 253 & M82 (NGC 3034)*

NGC 253 and M82 are amongst the brightest IR galaxies and are considered the prototypical starbursts. At ~2.6 Mpc NGC 253 is also one of the closest. The OSSE spectrometer aboard CGRO has already positively detected continuum emission from NGC 253 up to ~165 keV (Bhattacharya et al., 1994) and very recently the CANGAROO Cherenkov telescope collaboration has also reported a detection of this source in TeV-range $\gamma$-rays (Itoh et al., 2003a). Although those authors have favored an explanation in terms of electronic IC emission (Itoh et al., 2003b), others have provided an alternative theory: that the high energy emission may be due to nucleonic cosmic rays accelerated to high energies via the collective effects engendered by the starburst activity (Romero & Torres, 2003; see also Völk, 2003; and Sugai et al., 2003). In effect, these starbursts can be considered simply scaled-up versions of Galactic HII regions mentioned in the previous section. The goal of further multiwavelength (in particular, $\gamma$-ray) observations would be to confirm the soft $\gamma$-ray emission from these starbursts and, in combination with future AGILE and GLAST datasets, to resolve the question over the type of particles producing the emission.

At 3.2 Mpc M82 is only slightly further away than NGC 253, and this alone may have been the cause of its non-detection in the OSSE datasets *until 1994* (Battacharya et al., 1994). Note that M82 was detected by the HEAO A4 detector (Gruber & MacDonald, 1993), and future multiwavelength studies are suggested.

## REFERENCES

Aharonian et al., 2001 A&A, 370, 112
Aharonian et al., 2002A&A, 393L, 37
Bhattacharya et al. 1994, ApJ, 437, 173
Butt et al., 2001ApJ, 562L, 167
Butt et al., 2002 Nature 418, 499
Butt et al., 2003, ApJ *accepted*, astro-ph/0302342
Bykov, 2001 SSRv, 99, 317
Borkowski et al., 2001 ApJ, 550, 334
Cesarsky & Montmerle 1983 SSRv, 36, 173
Drury et al., 2001 SSRv, 99, 329
Dunne et al., 2003 ApJ, 590, 306
Enomoto et al., 2002 Nature 416, 823
Erlykin & Wolfendale, 1997, J. Phys. G, 23, 979
Erlykin & Wolfendale, 2003, J. Phys. G, 29, 709
Gruber & MacDonald, 1993 *priv. comm. as plotted in Figure 2* of Bhattacharya et al., 1994
Gvaramadze & Vikhlinin, 2003A&A, 401, 625
Hanson, M, 2003, ApJ in press, astro-ph/0307540
Hartman et al., 1999ApJS, 123, 79
Itoh et al., 2003a A&A, 402, 443I
Itoh et al., 2003b, ApJ Lett, submitted, astro-ph/0301147
Kanbach et al., 2003 (SPIE) *X-Ray and Gamma-Ray Telescopes and Instruments for Astronomy*. Edited by Joachim E. Truemper, Harvey D. Tananbaum. Proceedings of the SPIE, Volume 4851, pp. 1209-1220
Knödlseder, 2003 proceedings of the IAU Symp 212, A Massive Star Odessey from Main Sequence to Supernova, *in press*
Moffat et al., 2002 ApJ, 573, 191
Plaga 2002NewA....7..317P

Parizot, 2002, in the Proceedings of the XXXVIIth RENCONTRES DE MORIOND meeting, Les Arcs, March 9-16, 2002.
Pfeffermann & Aschenbach, 1996 *Proc. 'Röntgenstrahlung from the Universe'*, eds. Zimmermann, H.U.; Trümper, J.; and Yorke, H.; MPE Report 263, p. 267-268
Plaga, R., 2002 New Astronomy, Volume 7, Issue 6, p. 317
Reimer, O., & Pohl, M., 2002A&A, 390L, 43
Romero & Torres, 2003ApJ, 586L, 33
Rosado et al., 1996A&A, 315, 243
Schlickeiser et al., 1982, A&A, 106L, 5
Shklovskii, I. S., Dokl. Akad. Nauk SSSR, 91, No 3, 475-478 (1953) [Library of Congress Translation # RT-1495]
Slane et al., 1999ApJ, 525, 357S
Sugai et al., 2003ApJ, 584L, 9
The et al., *proc. of the 4th COMPTON symposium*, April 27-30, 1997; astro-ph/9707086
Thorsett et al., 2003, 2003ApJ, 592L, 71
Torres et al., 2003, Physics Reports, Vol. 382, No. 6, August 2003
Townsley et al.,2003, ApJ in press, astro-ph/0305133
Vink et al., 2000A&A, 362, 711
Uchiyama et al., 2002 PASJ, 54L, 73
Völk, 2003 *Invited paper presented at "The Universe Viewed in Gamma-Rays"* held on 25-28 Sept. 2002 in Kashiwa Japan , astro-ph/0303078
Watanabe et al., 2003, *Proceedings of the 28th ICRC*, pp. 2397-2400, in press
Westerlund, 1969, AJ, 74, 879
Wilson et al., 1970, A&A, 6, 364
Wolk et al., 2002ApJ, 580L, 161

This brief review benefitted from discussions with A. Bykov, P. Benaglia, J. Combi, T. Dame, J. Drake, Ph. Durouchoux, M. Kaufman Bernadó, P. Milne, F. Miniati, M. Pohl, O. Reimer, J. Rodriguez, G. Romero, M. Rupen, D. Torres, and S. Wolk. The support of the CHANDRA project, NASA Contract NAS8-39073, is gratefully acknowledged.